\documentclass[aps,pre,superscriptaddress,twocolumn,showpacs,floatfix]{revtex4}

\usepackage{graphicx}
\usepackage{dcolumn}
\usepackage{bm}

\usepackage{epsfig}
\usepackage{amsmath}

\newcommand{\beq}{\begin{equation}}
\newcommand{\eeq}{\end{equation}}

\newcommand{\bea}{\begin{eqnarray}}
\newcommand{\eea}{\end{eqnarray}}

\begin{document}

\title{Record dynamics and the observed temperature plateau in
the magnetic creep rate of type II superconductors.}
\author{L.P. Oliveira} 
\author{Henrik Jeldtoft Jensen}
\email[Author to whom correspondence should be addressed:\\]{h.jensen@ic.ac.uk}
\homepage{http://www.ma.ic.ac.uk/~hjjens/}
\affiliation{Department of Mathematics, Imperial College London,
South Kensington campus, London SW7 2AZ, U.K.}
\author{Mario Nicodemi}
\affiliation{Universit\'a di Napoli ``Federico II'',
Dip. Scienze Fisiche, INFM and INFN, Via Cintia, 80126 Napoli, Italy }
\author{Paolo Sibani}
\affiliation{Fysisk Institut, Syddansk Universitet, 5230 Odense M, Denmark}

\date{\today}

\begin{abstract}
We use Monte Carlo simulations of a coarse-grained three dimensional model 
to demonstrate that the experimentally observed approximate temperature 
independence of the magnetic creep rate for a broad range of temperatures
may be explained in terms of record dynamics, {\it viz.} the dynamical
properties of the times at which a stochastic fluctuating signal 
establishes records.
\end{abstract} 


\maketitle

\section{Introduction} 
The magnetization of type II superconductors is determined by the number
of quantized magnetic vortices inside the sample \cite{tinkham}. 
As an externally imposed  magnetic   field is increased,  vortices penetrate the sample
in a process which at non-zero temperature  is driven by thermal activation
over energy barriers produced by the sample surface and by pinning centers 
in the bulk. When the external field is lowered, vortices   leak out
of the sample. The rate at  which   vortices   move in and 
out of the sample determines the magnetic creep rate.

Given that the magnetic relaxation is driven by thermal activation, it
is rather surprising that experiments have found the creep rate
to be essentially temperature independent in a wide temperature range 
\cite{yeshurun,mota,fruchter,stein,aupke,mota_org,mota_hf,pollini_1,pollini_2,hoekstra,nowak}. 
Several  mechanisms have  been suggested to explain how competing factors are able to
cancel the typical Arrhenius temperature dependence $\exp(U/T)$ for activation
over an energy barrier $U$. The most prominent theoretical suggestion 
so far is probably the description in terms of collective creep \cite{Blatter}.

Here we show  that  the lack of temperature dependence of the
creep rate  can be understood  quite simply in terms 
of \emph{record dynamics}\cite{Sibani93a}, which
has recently been proposed as a general 
mechanism for the irreversible dynamics 
following a sudden quench in glassy systems~\cite{Sibani93a,Anderson04}.

We base our analysis on the following two experimental and numerical 
observations:
\begin{itemize}
\item[A)] For a long time glitches have been observed in the time dependences of the
magnetic relaxation of type II superconductors (see \cite{nowak}
and references therein). As the external magnetic field is varied,  the magnetization
undergoes abrupt jumps whenever vortices suddenly move in  or out  of
the sample. 
\item[B)] A number of studies, experimental as well as theoretical 
\cite{ghosh,roy,ravikumar,kokkaliaris,banerjee,jackson,nicodemi_1,nicodemi_2,nicodemi_3,nicodemi_4,nicodemi_5,nicodemi_6}, 
indicates that in a broad range of not too high temperatures the vortex system 
exhibits many characteristic features of glassy dynamics. 
\end{itemize}

Many~\cite{daeumling,chikumoto,yang,yeshurun_2,cai,zhukov,pradhan,kupfer}  
but far from  all~\cite{ghosh,roy,ravikumar,kokkaliaris,banerjee} creep experiments 
study the response of the external magnetic field to a sweep,  including sign 
reversal.  For simplicity, we will presently   concentrate on 
a setup in which the external field is initially ramped at a  fixed  rate up to
a value and  then remains   constant for the entire duration of the
experiment.

We  use a  three dimensional version of the Restricted Occupancy Model 
(ROM)\cite{jackson,nicodemi_1,nicodemi_2,nicodemi_3,nicodemi_4,nicodemi_5,nicodemi_6}
to study the response to a fixed applied magnetic field.
Important physical properties of the 3D layered ROM model have a nearly 
temperature independent 
dynamical evolution, matching in this respect  simulation results
previously obtained for the same model
\cite{jackson,nicodemi_1,nicodemi_2,nicodemi_3,nicodemi_4,nicodemi_5,nicodemi_6} 
as well as experimental results on type II superconductors
\cite{mota,fruchter,stein,aupke,hoekstra,nowak,ghosh,roy,ravikumar,kokkaliaris,banerjee,daeumling,chikumoto,yang,yeshurun_2,cai,zhukov,pradhan,kupfer}.
As a strong temperature sensitivity is expected for the activated
dynamics of an entirely  classical model system, the mechanism behind
the observed temperature independence warrants some further theoretical scrutiny.

A fixed applied magnetic field  can be expected to 
lead  to a non-stationary time dependence of the internal  magnetization  
which slowly increases   from zero to about the value of the applied field. 
Interestingly,   the low temperature dynamical evolution of the response involves two 
types of configuration rearrangements, having widely different time scales. 
One type consists of rapid glitches by which the magnetization irreversibly 
jumps to higher values, as additional vortices enter the system. We refer to
these glitches as {\em quakes} \cite{Anderson04,Sibani03,Dall03} to emphasize 
their non-equilibrium nature, their abruptness and their dramatic effect on the
state of the vortex system. The quakes are separated by much longer periods of apparent 
quiescence, during which the vortex system is `searching' for a
configuration of larger stability and for a way to accommodate more vortices
inside the sample. Even though the total magnetization does not change, 
the internal spatial organization of the vortices undergoes considerable rearrangements.  

We argue that  the slow vortex creep associated with the quakes 
can be analyzed through the statistical properties  of  record  
dynamics, which   immediately explains the temperature 
independence of the creep rate\cite{Sibani93a} for the range of temperatures for which this
description is applicable. By record dynamics we mean the following. Consider
a stochastic time signal $\chi(t)$. The corresponding record   signal is
$R(t)=\max_{\tau<t}\{\chi(\tau)\}$, i.e.  the largest value assumed by
$\chi(t)$ up to the current time $t$. The physical 
idea behind record dynamics~\cite{Sibani93a,Sibani03,Sibani04,Sibani04a}
is that   large irreversible configurational changes in noisy systems with 
a macroscopic number of metastable attractors are induced by noise fluctuations
of record size. Similar behavior is observed in other glassy 
metastable systems, e.g.\ gels and spin-glasses, and can be characterized 
statistically in a similar way \cite{Anderson04,Sibani03,Dall03}.  
This highlights  the underlying unity of 
non-equilibrium glassy dynamics at low temperatures and 
supports  the possibility of a common theoretical description.

The paper is organized as follows: The next Section summarizes the properties
of the ROM model used in the simulations. Section~\ref{T-independence}
briefly introduces  a possible mechanism~\cite{Sibani04,Sibani04a,Anderson04}
by which  activated dynamics can become insensitive to the temperature in 
a glassy system, and demonstrates its relevance for the ROM model dynamics. 
Section~\ref{results} focuses on creep rates and contains comparison
with experiments.
Finally, Section~\ref{summary} presents 
a summary and a discussion. 

\section{The 3D ROM model}
\label{model}
The two dimensional  Random Occupancy Model (ROM)
has been shown  to reproduce the essential features of vortex 
dynamics at nonzero temperature 
\cite{jackson,nicodemi_1,nicodemi_2,nicodemi_3,nicodemi_4,nicodemi_5,nicodemi_6}.
Here we use Monte Carlo (MC) simulations of a generalized  
three dimensional layered version of the ROM model  
to capture the long time relaxation of interacting vortex matter.

In vortex matter, the length scale of the interactions can be very large
compared with the average distance between (pancake) vortices. At high
densities, this implies that each vortex interacts with many others. 
For layered superconductors this situation can roughly be described 
by two length scales: the first is the range of the interaction parallel to
the planes, this is the London
penetration depth $\lambda$. The second length
scale is the vortex correlation length, $\xi_{||}$, parallel to the
applied field (which we imagine to be perpendicular to the   
copper oxide planes for high temperature superconductors).
The exact identification of this length scale is difficult
and is likely to depend on the anisotropy of the material,
the nature of the pinning, the strength of the magnetic induction
and on the temperature. This length scale may be related to vortex
line 
cutting\cite{Puig00,Goffman98,delaCruzBSCO,HenrikYBC0,OdeficientYBC0,Gaifullin,Busch,Fuchs}.  
These length scales  respectively give the horizontal
and vertical lattice spacing of our model. The horizontal 
coarse-grained length scale $l_0$, corresponds to the penetration depth 
$l_0=\lambda$ of the superconducting material, and the 
spacing between the layers in our lattice we consider 
$l_1\sim \xi_{||}$. Smaller length scales are ignored. 
For our purposes this approximation is acceptable because the length 
scales smaller than $\lambda$ seem to have little influence on the 
long time glassy properties of vortex matter.

Another limitation of the model is that it ignores the 
variation of $\lambda$ with the temperature. As will
become clear from our ensuing discussion of record dynamics, 
ignoring the temperature dependence of $\lambda$ is not crucial
for our explanation of the observed temperature plateau of
the creep rate. 

In a sample of a superconducting material the vortex matter behavior is
determined by the competition of four energy scales \cite{Blatter}: intra
and interlayer vortex-vortex interaction, vortex-pinning interaction and
thermal fluctuations, all of which are schematically included in the ROM
model.\bigskip

The Hamiltonian of the ROM model is thus the following:
\begin{equation}
H=\sum_{ij}A_{ij}n_{i}n_{j}- \sum_{i}A_{ii}n_{i}+\sum_{i}A_{i}^{p} n_{i} +
\sum_{\left\langle ij\right\rangle _{z}}A_{2}\left(  n_{i}-n_{j}\right)  ^{2},
\label{hamilton}
\end{equation}
where $n_{i}$ is the number of vortices on site $i$ of the lattice. 
In a superconducting sample the number of vortex lines per unit area is
restricted by the upper critical field ($B_{c2}$) \cite{tinkham}, so in the
model the number of vortices per cell can only assume values smaller than
$N_{c2}=B_{c2}l_{0}^{2}/\phi_{0}$ \cite{Nicodemi_and_Henrik_c1,nicodemi_4}.
Hence the name Restricted Occupancy Model. Moreover,
as we are interested in
a simulation setup that does not require magnetic field inversion and the
vortex-antivortex creation is strongly suppressed, 
we simply consider $n_{i}\geq0$.

The first two terms in Eq. (\ref{hamilton}) 
represent the repulsion energy due to vortex-vortex
interaction in the same layer, and the vortex self energy respectively. Since 
the potential that mediates this interaction decays exponentially at
distances longer than our coarse-graining length $\lambda$, interactions 
beyond nearest neighbors are neglected. We set
$A_{ii}:=A_{0}=1$, $A_{ij}:=A_{1}$ if $i$ and $j$ are nearest neighbors on
the same layer, and $A_{ij}:=0$ otherwise.

The third term represents the interaction of the vortex pancakes with the pinning
centers. $A_i^{p}$ is a random potential and for the purposes of this work we
consider that $A_i^{p}$ has the following distribution $P\left(  A_i^{p}\right)
=\left(  1-p\right)  \delta\left(  A_i^{p}\right)  -p\delta\left(  A_i^{p}
-A_{0}^{p}\right)$.
The pinning strength $\left|  A_{0}^{p}\right|$ represents the total
action of the pinning centers located on a site. In the present work
we use $\left|  A_{0}^{p}\right|  =0.3$.

Finally the last term describes the interactions between the vortex sections
in different layers. This term is a nearest neighbor quadratic interaction
along the $z$ axis, so that the number of vortices in neighboring cells along
the $z$ direction tends to be the same.

The parameters of the model are defined in units of $A_{0}$. The
time is measured in units of full MC 
sweeps. The relationship between the model parameters and material parameters is
discussed in \cite{Nicodemi_and_Henrik_c1,nicodemi_4}.

Each individual MC update involves the movement to a neighbor site
of a single randomly selected vortex. The movement of the vortex is
automatically accepted if the energy of the system decreases; if the energy of
the system increases, the movement is accepted with probability 
$\exp(-\Delta E/T)$ \cite{Binder}. 

Given that the movement of pancake vortices is restricted to 
the superconducting planes we only allow MC movements parallel
to the planes. We have used periodic boundary conditions along 
the $z$ direction. 

The external magnetic field is modeled by the edge sites on each
of the planes. The density at the edge is kept at a controlled value. 
During a MC sweep vortices may move between the bulk sites and the edge sites.
After each MC sweep the density on the edge sites is brought 
to the desired value. Initially the external field is increased to
a desired value ($N_{ext}=10$ vortices per edge site) by a very rapid 
increase in the  density on the edge sites.  
We have here used a sweeping rate $\gamma$ of 0.25
per MC sweep (compared with $\gamma\in[10^{-6},10^{-2}]$ in our
previous studies 
\cite{jackson,nicodemi_1,nicodemi_2,nicodemi_3,nicodemi_4,nicodemi_5,nicodemi_6}).
After this fast initial ramping the external field is kept constant, while
we study how the vortices move into the sample. The age of the system, $t_w$,
is taken to be the time since the initial ramping. 

We have studied systems of different sizes, 
and obtain similar results except for very
small system sizes. Our key
results were obtained in a system consisting of 8 layers
of size $16\times16$. The model parameters used in our simulations 
were:
\begin {align*}
\text {number of realization} &=4000 \\
A_{1} &= 0.28 \\
A_{2} &= 0.5 \\
p &= 0.5 \\
A_0^p &= 0.3 \\
N_{c2} &= 27 \\
\gamma &= 0.25 \\
N_{ext} &= 10 
\end {align*}

\section{Record dynamics}
\label{T-independence} 
We will in this section describe how the observed plateau in the temperature
dependence of the magnetic creep rate can be explained in the framework
of record dynamics. Let us first mention the salient features of what we
mean by record dynamics. Consider a stochastic signal $\chi(t)$ with
no time correlations. Now derive the
record signal $R(t)=\max_{\tau<t}\{\chi(\tau)\}$. We note that $R(t)$ is a
monotonous piecewise constant function which only  increases  its value
at discrete times $t_k$,  whenever  $\chi(t)$ manages to fluctuate to a 
value larger 
than any encountered previously. For our present purpose,  the most important
property  of the statistics of the record times $t_k$ is that  the probability
that exactly $q$ records occur during 
the time interval $[t_w,t_w+t]$ (where $t_w$ is the time since the initiation), 
is to a good approximation Poisson
 distributed on
a logarithmic time scale \cite{Sibani93a,Sibani03}, i.e.,
\begin{equation}
p(q)= {\langle q\rangle^q\over q!}\exp(-\langle q\rangle)
\label{log_Poisson_pdf}
\end{equation}
with average number of quake events proportional to logarithmic time 
\begin{equation}
\langle q\rangle= \alpha\log(1+t/t_w).
\label{ave_rate}
\end{equation} 
Here $\alpha$ is the logarithmic rate of events.
To get the gist of the mathematics behind Eq. (\ref{log_Poisson_pdf})
(for full detail see ref.~\cite{Sibani93a}),  
we note that  since the largest outcome, {\it i.e.} the record, of $t$ 
independent trials is equally
likely to occur at any of the $t$ instances in the sequence, it occurs at  the 
first attempt  with probability $1/t$. Hence, the probability
$p(1)$ of exactly one record in $t$ trials  is $1/t$,  
independently  of the distribution of the {\em  underlying signal} $\chi(t)$.  
It is important to point out that the same also holds
for the general expression for $p(q)$ in Eq. (\ref{log_Poisson_pdf}). 
The independence of $p(q)$ on the distribution
of random numbers corresponds to the
independence of    $\chi(t)$ on the thermal noise
and  will translate directly into
the temperature independence of the creep rate. 

It has recently been shown~\cite{Sibani04,Sibani04a,Anderson04} 
that in (glassy) systems having  a large number of dynamically inequivalent
attractors,  temperature independence  of suitably coarse-grained dynamical 
variables can arise from the peculiar way in which the  attractors are
selected  as the system evolves from a typically rather unstable initial
configuration through gradually more stable ones. A similar  noise insensitivity 
of stochastic dynamics has been observed with other types of noise, 
{\it  e.g.}\ in driven dynamical systems \cite{Sibani93a,Sibani01} 
and in evolutionary dynamics\cite{Sibani99a,Anderson04}.
\begin{figure}
\includegraphics*[angle=0,width=8cm]{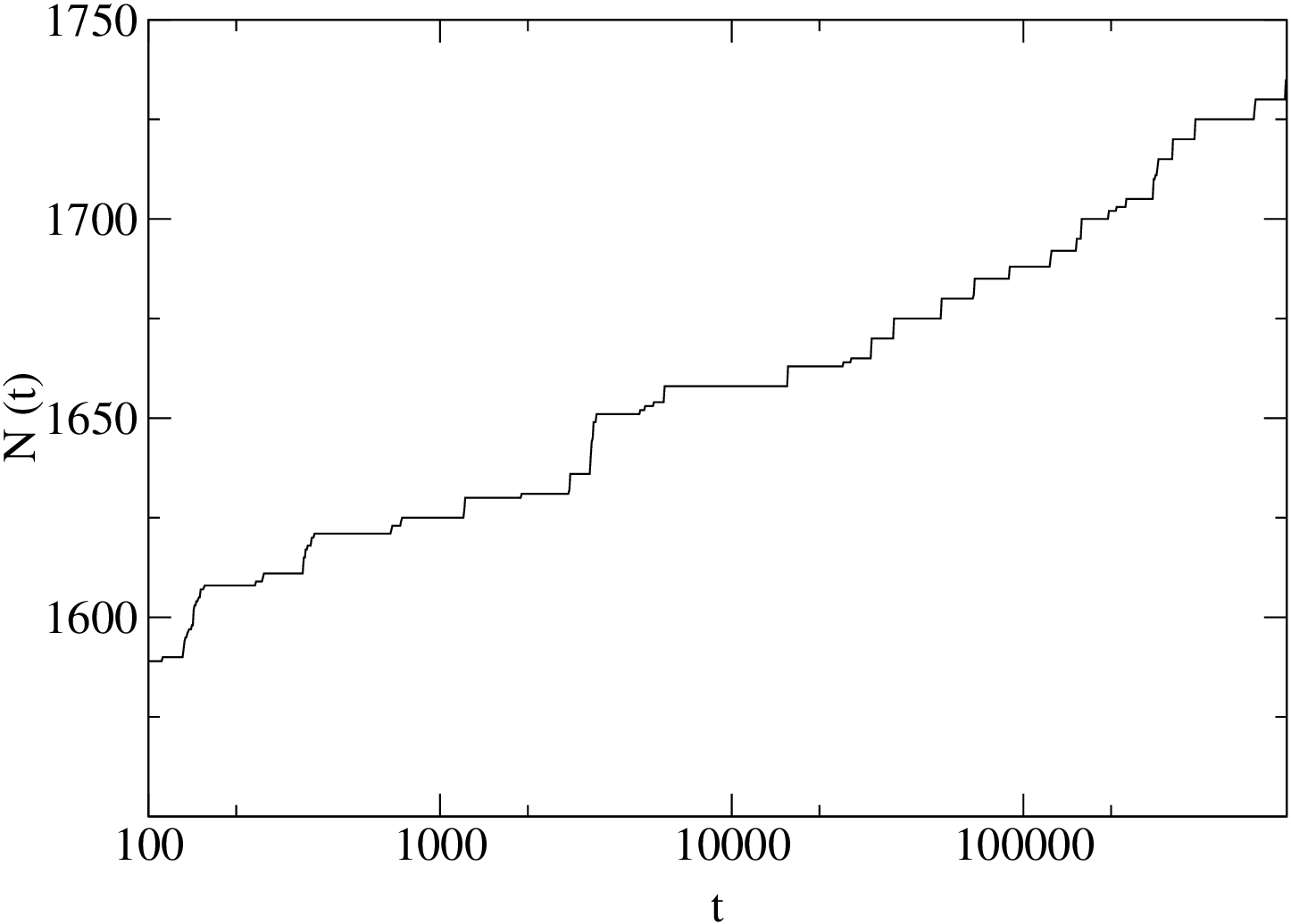}
\caption{The detailed time variation of the
total number of vortices $N(t)$ on the system for a
single realization of the pinning potential and the thermal noise in
a $8\times8\times8$ lattice for $T = 0.1$. Notice the monotonous
step function character of the time series.
\label{steps}
}
\end{figure}

That the ROM model  also  exemplifies this type of behavior can be gleaned
from Fig.~\ref{steps}, showing  the time dependence 
of the number of vortices $N(t)$ which  during a single run  
have entered the system up to  time $t$. 
Importantly, the length of the 
quiescent periods typically increases with time -- notice the 
logarithmic time axis in Fig. \ref{steps}. Were this not the case, 
the dynamics would  appear continuous in terms of a suitably 
coarse-grained time scale. Conversely, the  lengthening of the 
intervals between successive quakes signals  the
anticipated  gradual entrenching of the dynamics  into
dynamically more stable configurations.

Also important is  that the overwhelming majority of the
observed glitches lead to states with a higher number of vortices. 
This de-facto irreversibility of the dynamics  enables one  to 
meaningfully approximate 
the  signal $N(t)$ with the monotonically increasing
record signal $R(t) = \max_{[0,t]}\{N(t)\}$.
We stress that this theoretically convenient idealization is only    
applicable  within the strongly non-equilibrium regime of our 
present concern. 
\begin{figure}
\includegraphics*[angle=0,width=8cm]{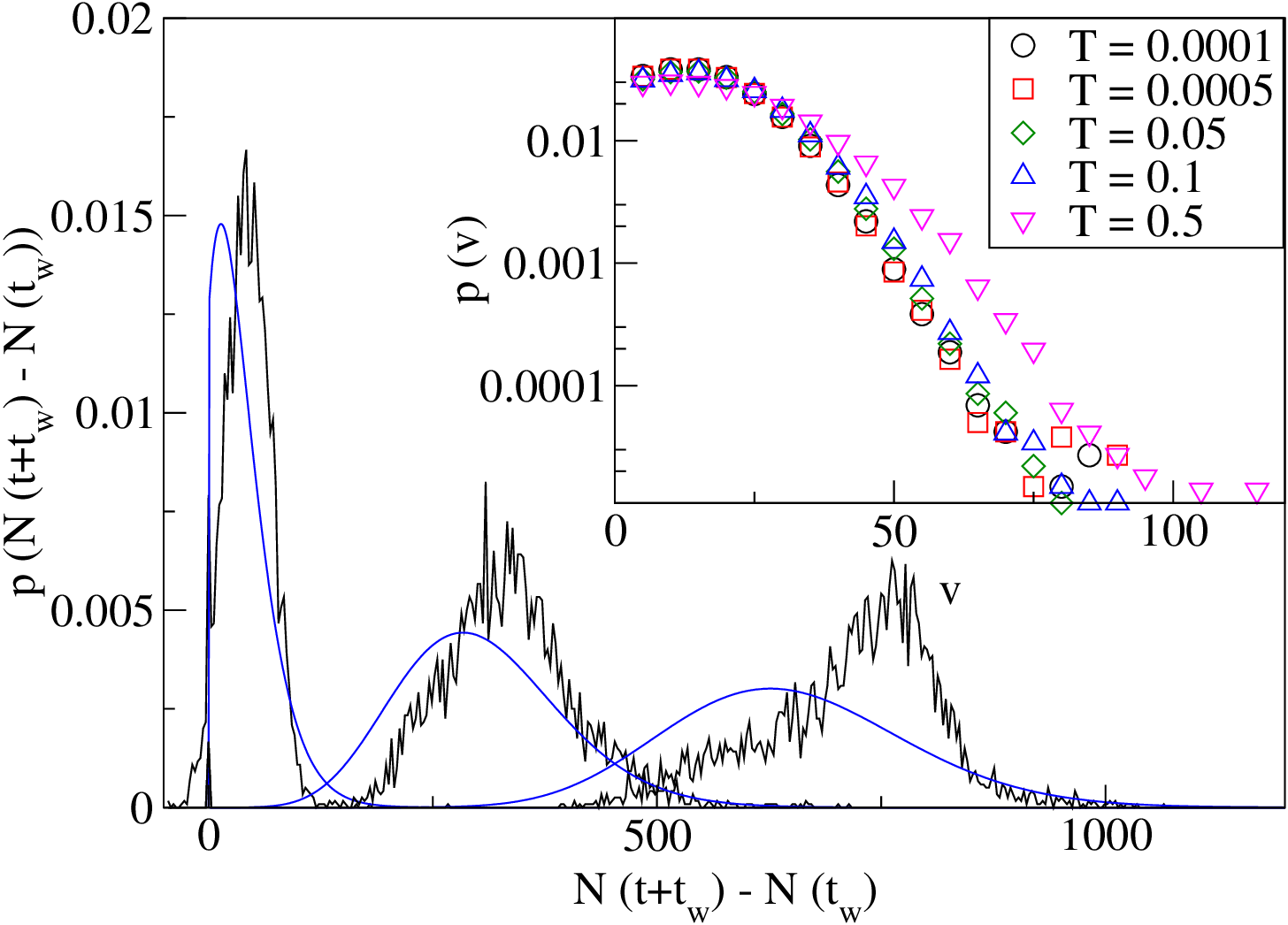}
\caption{(Color online) The main panel contains
the temporal evolution of the probability
density function, $P(N(t+t_w)-N(t_w))$, of the number of vortices entering
for $t_w=1000$ and three different observation times 
$t= 188, 2791, 8371$ given by the black, slightly jagged, curves.
The smooth curves (blue) are a
fit to the theoretical expression  (see Eq. (\ref{pdf_N})).
The system is $16\times16\times8$ and $T=0.0001$.
The insert shows the quake size distribution for
various temperatures for the time interval
between $t=1000$ and $t=10000$. For $T \le 0.1$ the distribution
has an approximately exponential tail. For $T=0.5$ the shape gets
closer to a Gaussian.}
\label{log-Poisson}
\end{figure}

Statistical insight into the time evolution
in the number of vortices present within  the system is provided  
by Fig.~\ref{log-Poisson}, where  the  empirical distributions of $N(t)$ 
are  displayed for three  different times, which are equidistantly 
placed  on a logarithmic time axis.
The insert in Fig. \ref{log-Poisson} shows  the tail of the
probability density function (pdf) of the number of vortices, $p(v)$, 
entering during a   single  quake. To a good approximation,  the tail 
is exponential and the  observed time dependence
and temperature dependence of  of $p(v)$  are negligible, except for the
highest temperature $T=0.5$. 

The interpretation
in terms of record dynamics suggests that the probability
that exactly $q$ quakes occur during the time interval $[t_w,t_w+t]$ 
is Poisson distributed on
a logarithmic time scale \cite{Sibani93a,Sibani03} 
according to Eq. (\ref{log_Poisson_pdf}).
We approximate the pdf for the number of vortices $v$ which enter
during a given quake event (see insert Fig. \ref{log-Poisson}) by an
exponential distribution $p(v)=exp(-v/\bar{v})/\bar{v}$, and assume
that subsequent quakes are statistically
independent.  The number of vortices entering during exactly $q$
quakes is then a sum of exponentially distributed
independent variables, and is hence  Gamma distributed.
We finally obtain the pdf for the 
total number of vortices   entering  during $[t_w,t_w+t]$ by 
averaging the  Gamma distribution for $q$ quakes
over the probability Eq. (\ref{log_Poisson_pdf}) that precisely
$q$ quakes  occur within the time interval of interest.
This leads to the following expression for 
the pdf of total number of vortices $\Delta N = N(t+t_w)-N(t_w)$
which may have entered  during the time interval $[t_w,t_w+t]$
\begin{equation}
p(\Delta N,t)= e^{-{\Delta N\over \bar{v}}-
\langle q\rangle} \sqrt{{\langle q\rangle\over \bar{v}\Delta N}}
I_{1}\biggl(2\sqrt{{\langle q\rangle\Delta N\over \bar{v}}}\biggr),
\label{pdf_N}
\end{equation}
where $I_{1}$ denotes the  modified Bessel function of 
order $1$\cite{Abramowitz}.
The above theoretical prediction, is compared  in Fig. \ref{log-Poisson}    
with our simulation results. 
To estimate $\langle q \rangle$ according to Eq. (\ref{ave_rate}), we
used $\alpha = 22.6$, as obtained from 
the logarithmic rate of the quake events. We can determine the 
average $\bar{v}$ in two ways. Either directly from the simulated
distributions in the insert of Fig. \ref{log-Poisson} or 
from fitting Eq. (\ref{pdf_N}) to the simulated data in the
main frame of Fig. \ref{log-Poisson}. In both cases we 
find $\bar{v} = 16$. We also find that $\bar{v}$ is essentially
temperature independent for temperatures below $T\approx0.1$.
This is also expected from the insert in Fig. \ref{log-Poisson}.
It is important to mention that the MC dynamics does overcome
plenty of positive energy barriers, $\Delta E>0$, through 
thermal activation for 
temperatures in the range $0.01<T<0.1$. As the temperature is
lowered fewer MC updates correspond to $\Delta E>0$ and
for the lowest temperatures MC steps involve $\Delta E\le0$ 
only\cite{nicodemi_3}. Nevertheless, the record dynamics remains
essentially temperature independent for $T<0.1$.

The agreement is encouraging and suggests that the process of vortex 
penetration into the sample can be described in 
terms of a Poisson process with logarithmic time argument, 
for short the  log-Poisson process. 
We also note  that the log-Poisson statistics
covers  the  temporal distribution
of the quakes but  has nothing to say on the  size distribution
of the jumps, {\it i.e.}\ the number $v$ of  vortices entering   
during a single event. This  stochastic quantity could in principle  
introduce additional  time and temperature dependencies. However, as  
mentioned, the insert of Fig. \ref{log-Poisson} shows that for a
very broad parameter range this is not the case. Accordingly, the creep rate
obtained by convoluting the distribution of the quake sizes and the 
log-Poisson distribution of number of quakes
will also be temperature independent.

The link between record statistics and  the
stochastic dynamics of a glassy   system is discussed
in detail in ref.\cite{Sibani03} on the  basis of 
several idealized physical assumptions.  
The first  element is the existence of a large number, in principle
a continuum,  of attractors. These 
are sets of configurations clustered around a local 
energy minimum and supporting equilibrium-like reversible thermal 
fluctuations. By contrast,  attractor changes---our
quakes---are assumed to be irreversible on the time scale
at which they occur. The exact nature of the quakes is 
not entirely clear in our system. At intermediate temperatures they are
related to activation over barriers and the jump in $N(t)$ is
associated with a increase in the energy of the interacting vortices.
However, at the lowest temperatures there is not
sufficient thermal energy available for the system to climb
over energy barriers. The thermal fluctuations are only able to 
push the vortices along
equipotential trajectories or to lower potential energy configurations. 
In this regime the vortex motion is hindered by jamming and the 
quakes are of a 
mechanical nature \cite{nicodemi_3}.

An interpretation in terms of record dynamics implies that the dynamical 
bottlenecks overcome by fluctuations are determined by the 
actual noise history,  and not  predetermined in a static fashion.
For other model systems~\cite{Sibani93a,Sibani01}, 
the validity of the assumed linkage  between noise records and barriers
was confirmed by considering white noise perturbations drawn from
a distribution with finite support, {\it e.g.}\ a box distribution,    
and by then  studying the properties of the selected 
attractors as a function of the maximum size of noise. 

Record-induced dynamics has thus a number of testable predictions,  
the most interesting of which is, for our purposes, 
the logarithmic time dependence and the 
striking temperature independence of the number of quakes occurring in 
the time interval $[0,t]$, which are shown in  the following Section. 

 \section{\bigskip Creep rates}
\label{results}
\bigskip

Let us now turn to the dependence on time and temperature of the 
total number of vortices in the sample.  At time $t=0$ we rapidly
increases the external field from zero up the value $N_{ext}$
(see section II). 
The vortex density of the bulk sites gradually increases 
as vortices move in from the boundary.

\begin{figure}
[ptb]
\includegraphics*[angle=0,width=8cm]{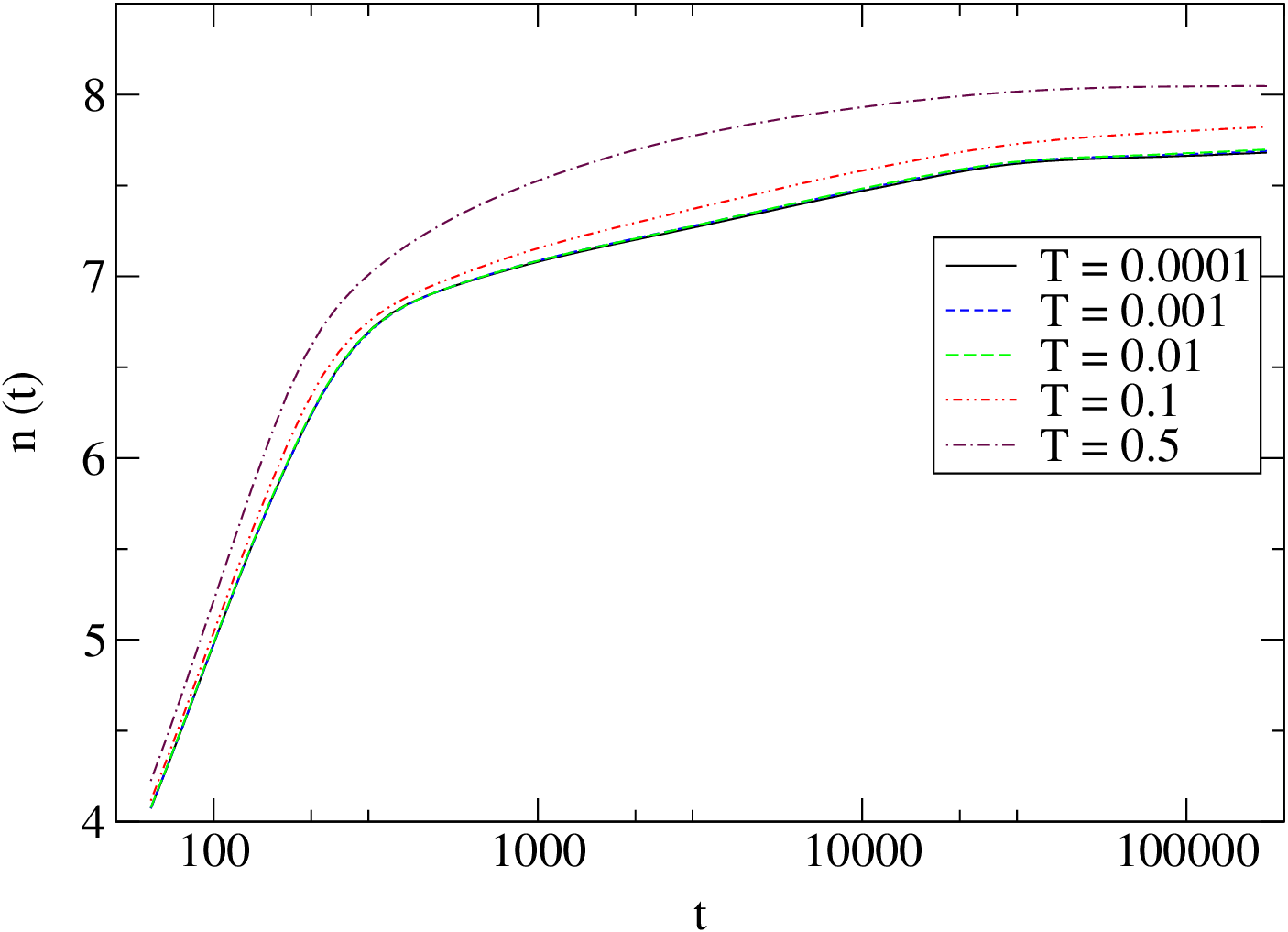}
\caption{(Color online) Vortex density time dependence in the system for
various temperatures. For
$T \le 0.1$ the vortex number
is a piecewise linear function of $\log(t)$. For $T = 0.5$
the system relaxation becomes faster.
In this plot $n(t)$ is averaged over many system realizations.
}
\label{N(log(t))}
\end{figure}

In Figure \ref{N(log(t))} we present, for a very broad range of 
temperatures, the average density $n(t)$ of the bulk sites 
as function of the natural logarithm of time $\log(t)$. As anticipated, 
the time dependence is temperature independent for all but the highest 
temperatures.  

One can identify three different 
temporal regimes separated at times $t_1\approx300$ and 
$t_2\approx 3\times 10^4$. 
For the remaining of this paper we will focus our
analysis on the intermediate regime $t_1<t<t_2$ (and choose $t_w=t_1$). 
Our reason for this is that at $t_1$ 
vortex interactions become essential through out the entire system.  
The late time regime $t>t_2$ is very difficult to resolve 
appropriately in simulations and probably equally difficult
to study experimentally.

For times $t_1<t<t_2$ Fig. \ref{N(log(t))} demonstrates that $N(t)$  
depends linearly on $log(t)$ to a very good approximation.  

The linear logarithmic time dependence is of course 
entirely consistent with the  record dynamics outlined in 
the previous section. We consider the total number of 
vortices in the system $N(t_{w}+t)$ to be the accumulated effect
of vortices entering during quake events that have occurred prior
to time $t_{w}+t$. Let $t_k$ denote the time of occurrence of quake number $k$
and let $v_k$ denote  the actual number of vortices entering during
this quake. We then have
\begin{equation}
N(t + t_{w}) = N (t_{w}) + \sum_{t_{w}<t_k<t_{w}+t} v_k,
\end{equation}
where the sum is over all quakes that occured during the time
interval $[t_w,t_w+t]$

From Fig. \ref{log_Poisson_pdf} we know that  $v_k$ is temperature
independent and possesses a well defined average $\bar{v}$. 
Since the  average number of quakes increases according to Eq. (\ref{ave_rate}),   
record dynamics predicts  the following (temperature
independent) temporal evolution of the average number of vortices  
\begin{equation}
\Delta N\equiv \langle N(t+t_{w})\rangle - \langle N (t_{w})\rangle 
=  \alpha \bar{v} \log(1+t/t_w).
\label{avN}
\end{equation}
{\it i.e.} for the considered time regime $t/t_w \gg 1$ a temperature 
independent logarithmic rate given by 
\begin{equation}
d\Delta N/d\log(t)= \alpha\bar{v}/(1+t_w/t) \approx \alpha\bar{v}.
\end{equation}

\begin{figure}
\includegraphics*[angle=0,width=8cm]{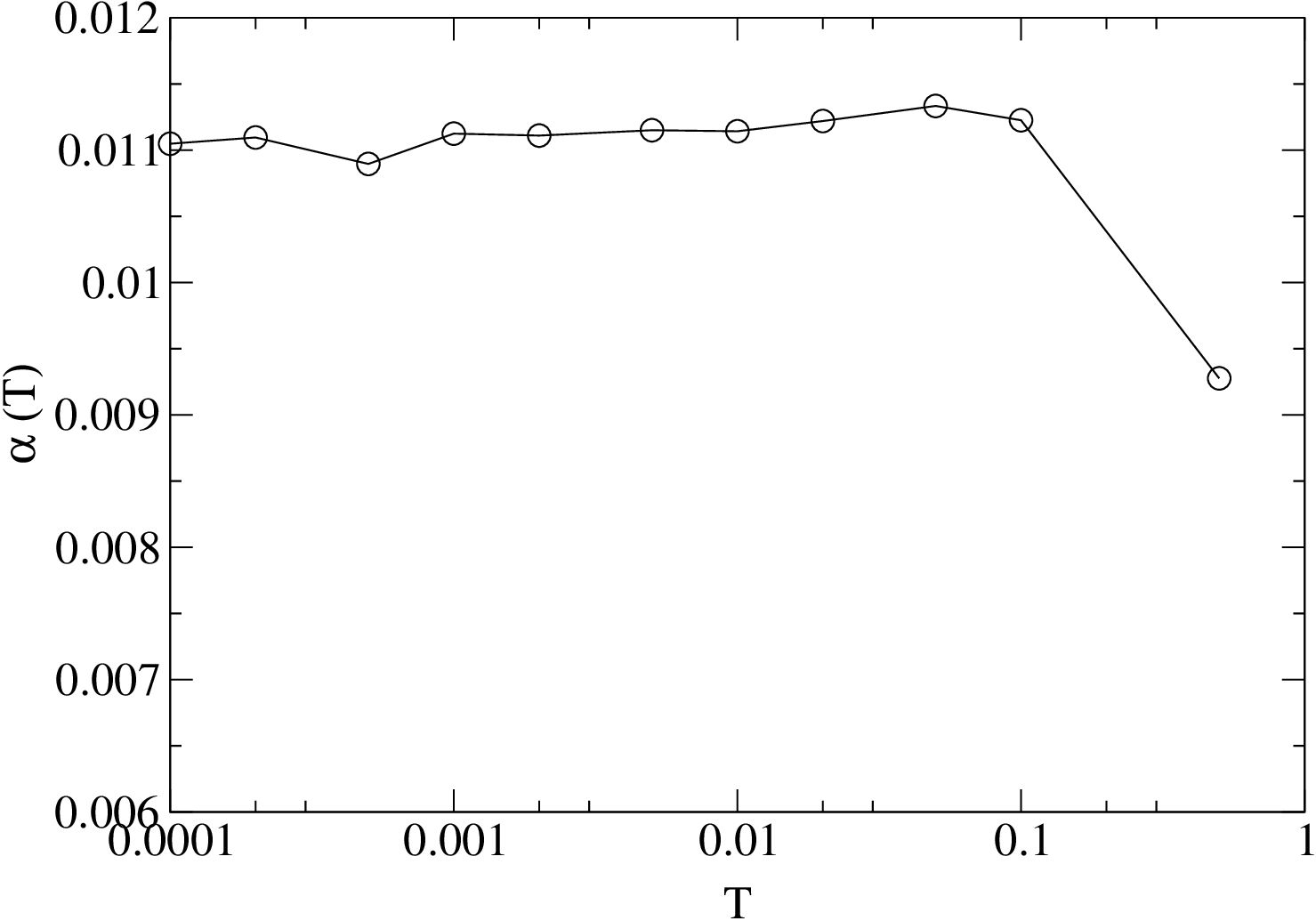}
\caption{Quake dynamics rate, $\alpha$, versus $T$ for the time
interval between
$t=1000$ and $t=10000$. Notice that $\alpha$ is roughly
temperature independent for three
orders of magnitude below $T<0.1$.
}
\label{alpha}
\end{figure}

We extract the rate of the quakes in the simulations from temporal 
signals like the one exhibited in  Fig. \ref{steps}. 
In Fig. \ref{alpha} we demonstrate that the quake rate 
is indeed approximately independent of temperature 
in the broad temperature interval
$10^{-4}<T<2\times 10^{-2}$.

\begin{figure}
\includegraphics*[angle=0,width=8cm]{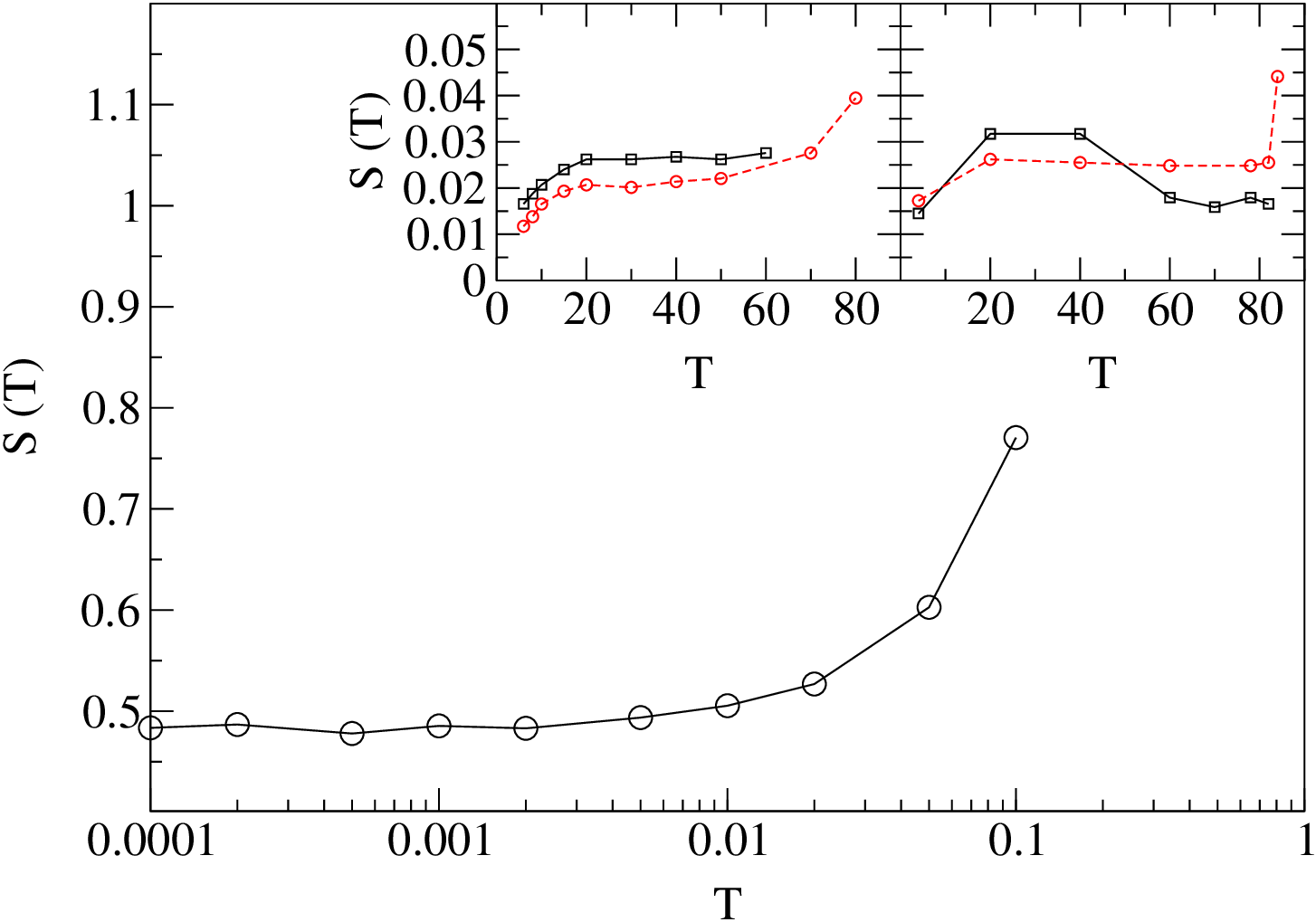}
\caption{(Color online) Main panel: Numerical results for the creep rate
versus
$T$ for the time
interval between $t=1000$ and $t=10000$. In the low temperature
region the creep rate is constant within our numerical precision for 
about two orders
of magnitude - we observe a nonzero creep rate in the
$T \rightarrow 0$ limit. 
Insets:
experimental
results for the creep rate versus T.
The right inset shows data from Keller {\it et al.} for melt
processed
YBCO crystals with the magnetic field applied along the $c$ axis (squares)
and $ab$ plane (circles). The left inset shows data from
\cite{Civale_et_al.} for unirradiated (squares) and 3 MeV proton-irradiated
(circles) YBCO
flux grown \cite{Kaiser_et_al.} crystals with a 1 T magnetic Field applied
parallel to the $c$ axis. 
}
\label{creep_rate}
\end{figure}

Finally, Fig. \ref{creep_rate} shows the near temperature independence
of the actual rate with which $N(t)$ changes. We extract this rate
from the data shown in Fig. \ref{N(log(t))} in the time region $t_1<t<t_2$ 
and plot the normalized creep rate $S=d \log[M(t)]/d\log(t)$ in order
to compare consistently with experiments, here we have used 
$M(t) = |N(t) - N_{ext}|$.

\section{Conclusion}
\label{summary}
We have presented an analysis of simulated vortex creep data in terms
of record dynamics. This approach allows us to interpret
the observed temperature independence of the creep rate
as a generic property of the dynamics of 
records obtained from the underlying fluctuating sequence. To
establish the temperature independence of the creep rate we do not need
to know the detailed nature of the quantity being gradually maximized.
Nor do we need a description of the intermittent vortex quakes which
are responsible for the abrupt changes in the number of vortices.
All we make use of is the assumption, supported by the simulated model,
that the abrupt glitches in the number of vortices inside the sample 
can be interpreted as arising from the records of some stochastic
process. We showed that the simulated creep rate behaves in a way 
very similar to published experimental data  for YBCO.

It is obvious that a better experimental and
theoretical understanding of the nature of the vortex quakes is interesting and
future study of the ROM model will  seek to improve our understanding
of the spatial and dynamical properties of these quakes. We can
already conclude that the physical mechanisms involved must be different
at low and high temperatures. At the lowest
temperatures activation over free energy barriers are excluded
and the quakes are related to mechanical rearrangements of the 
vortices\cite{nicodemi_3}. At elevated temperatures the quakes are 
expected to be triggered by activation over thermal barriers. 
The present paper shows that record dynamics can be used to understand
the temperature range from very low temperatures, where no barriers 
can be climbed, up through a regime where thermal activation does
take place. For high temperatures (in our case for $T>0.1$) the
description in terms of record dynamics breaks down. This happens
when there is sufficient thermal energy available to make
any trapped metastable configurations short lived.

Let us finally mention that our description in terms of record dynamics may
not exclude aspects of previous descriptions of vortex relaxation
in terms of {\em e.g.} correlated collective vortex creep \cite{Blatter}.
We would rather think of our approach as contributing to an
understanding of the detailed nature of the dynamics of the correlated
vortex regions. 

\section{Acknowledgments} We are indebted to Andy Thomas, Dan Moore and 
Gunnar Pruessner for their support with the computations. Support 
from EPSRC, the Portuguese FCT, a visiting fellowship from EPSRC 
and financial support from the Danish SNF are gratefully acknowledged. 


\end{document}